\documentclass{article}
\usepackage{amsmath} 
\usepackage{amsfonts, dsfont} 
\usepackage{authblk} 
\usepackage{hyperref} 
\usepackage{graphicx} 
\usepackage{apalike} 

\providecommand{\keywords}[1]{\textbf{\textit{Index terms---}} #1}

\title{End-to-end deep meta modelling to calibrate and optimize energy consumption and comfort}

\author{Max Cohen\footnote{This work was supported by grants from Région Ile-de-France}}
\author{Sylvain Le Corff}
\affil{Samovar, T\'el\'ecom SudParis, CITI, TIPIC, Institut Polyechnique de Paris}
\author{Maurice Charbit}
\author{Alain Champagne}
\author{Gilles Nozi\`ere}
\affil{Oze \'Energies}
\author{Marius Preda}
\affil{Samovar, T\'el\'ecom SudParis, ARTEMIS, ARMEDIA, Institut Polyechnique de Paris}
\date{}

\begin{document}
\maketitle
\begin{abstract}
	In this paper, we propose a new end-to-end methodology to optimize energy performance and thermal comfort in office buildings, without any renovation work. The process is decomposed into three steps: metamodel training, model calibration and optimization.
	We introduce and train a metamodel on thousands of weather and building settings scenarios, using samples from a physical simulation model.
	Its much faster computation time allows for the calibration of two weakly instrumented buildings, through a derivative free optimization procedure.
	Using historic data from these buildings, we estimate up to $60$ unknown parameters defined by energy managers, such as heat capacity, window area or exposition.
	Energy consumptions are finally minimized while maintaining a target thermal comfort using the Pareto front provided by a multi-objective optimization algorithm.
	Our approach allows the computation of the entire calibration-optimization pipeline on several types of buildings.
	Moreover, the numerical experiments illustrate how it may ensure a significant gain in energy efficiency, up to almost 10\%, while being computationally much more appealing than simulation programs.
\end{abstract}


\section{Introduction}
In 2009, the building stock accounted for over 40\% of the total French energy consumption, as well as almost a quarter of greenhouse emissions (Loi Grenelle\footnote{\href{https://www.legifrance.gouv.fr/loda/id/JORFTEXT000020949548/2020-09-21/}{Loi Grenelle I, Article 3}}).
In the ecological context where energy waste cannot be ignored, improving this energy efficiency is an important step. 
This translated in an objective of 38\% reduction in the consumption of the building stock for 2020, through the renovation of 400,000 apartments per year\footnote{\href{https://www.legifrance.gouv.fr/loda/id/JORFTEXT000020949548/2020-09-21/}{Loi Grenelle I, Article 5}}, raised to 500,000 six years later\footnote{\href{https://www.legifrance.gouv.fr/jorf/id/JORFTEXT000031044385/}{Loi relative à la Transition Energétique pour la Croissance Verte, Article 3}}.
However, despite setting higher and higher objectives, the actions carried out to date still fall short in terms of results, as stated by the Ademe (Agency for the environnement and energy)\footnote{\href{https://www.lefigaro.fr/conjoncture/accord-de-paris-pourquoi-les-pays-ne-sont-pas-a-la-hauteur-de-leurs-engagements-20190419}{Hervé Lefebvre, head of the Climat departiement of the Ademe, 2019}}.
According to the National Low-Carbon Strategy (SNBC), the average number of yearly renovations is expected to be around 370,000 for the period 2015-2030\footnote{\href{https://www.ecologie.gouv.fr/strategie-nationale-bas-carbone-snbc}{Stratégie Nationale Bas-Carbone (SNBC)}}.

The aim of this paper is to provide optimal building management settings governing Heating, Ventilation and Air-conditioning (HVAC) in order to improve thermal comfort and optimize energy consumption, without costly, invasive or time consuming renovation works.
Global energy demand for heating, ventilation and air-conditioning in commercial or public buildings has been increasing rapidly for the past few decades.
This rising demand is at the root of the complex problem of simultaneously maintaining a satisfactory thermal comfort in buildings and reducing the environmental impact.
This makes the analysis of building energy performance a challenging multi-criteria problem.

The solution proposed in this paper is decomposed into three steps: (i) designing a model to predict future energy loads and indoor temperature based on the HVAC system and weather data, (ii) calibrating the unknown parameters of this model with real data obtained from sensors in each building and (iii) optimizing the HVAC settings to minimize the total energy consumption in future periods while maintaining a given thermal comfort.
Step (i) can be achieved with simulators that describe heat transfers between the building and its environment such as EnergyPlus, TRNSYS or DOE-2.
They predict the indoor temperatures and consumptions from the description of the building, HVAC settings and weather data.
However, their accuracy relies heavily on the precision of the building description: window to wall ratio, exposition, thermal conductivity or heat capacity among others. Because our strategy does not involve renovation works on the building site and use a simplified building description we usually can only provide rough estimates of these input parameters.
The calibration procedure in step (ii) aims at estimating the parameters of the model proposed in (i) to provide accurate indoor temperatures and consumptions predictions.
We measure the discrepancy between model predictions and real data with a loss function that is minimized iteratively.
Because this minimization requires tremendous number of calls to the model designed in step (i), the calibration step quickly becomes excessively time consuming. 
A very popular practice in the building optimization literature consists in using surrogate models.

Metamodelling approaches substitute in step (i) the physical simulator with a much faster surrogate model to tackle calibration and optimization tasks. This surrogate model is trained on a dataset of simulations conducted by the simulator, that aims at exhaustively capturing building behaviors for various building geometries and management settings. The most recent statistical models make this step more and more relevant in our context, as their accuracy reaches new heights while their execution time is constantly reduced.
Yet, most approaches only implement the most simple architectures for building optimization.

In this paper, we first propose to train a metamodel based on Recurrent Neural Networks (RNN). We propose a comparison of several approaches which illustrates that sequence to sequence models, such as RNN, can yield a significant increase in performances with respect to the alternatives previously considered in our framework. In addition, we propose to compare such RNN to attention-based models which are the go-to architectures in other fields to predict sequences with complex dependencies. 
Our metamodel, which depends on a few physical parameters, is then calibrated using real data to provide accurate predictions for various buildings. Two real buildings were used to illustrate the flexibility of this approach. The final step of our end-to-end methodology consists in optimizing energy consumptions, while maintaining a given level of comfort. This multi-criteria optimization problem requires determining an optimal compromise between consumption and comfort, as improving one objective results in degrading the other. This optimization is achieved through an iterative algorithm, justifying our choice of replacing simulators with a faster surrogate model. Following this methodology, we were able to train and calibrate our metamodel and to reduce the hourly consumption of two buildings by 5\% and 10\%. 

\section{Related works}
\label{sec:related}
\subsection{Physical simulators}
Physical simulators based on thermal propagation equations are traditionally used to describe buildings.
The most common pieces of software available, EnergyPlus, TRNSYS or DOE-2, are used to simulate the system behavior based on a schematic view of the building.
EnergyPlus was used for instance in \cite{shabunko2018energyplus} to build three types of typical designs and to benchmark the energy performance of 400 residential buildings.
In \cite{zhao2016occupant}, the authors proposed a predictive control framework based on Matlab and EnergyPlus in order to optimize energy consumptions while meeting the individual thermal comfort preference.
In \cite{Magnier2010MultiobjectiveOO}, the authors highlighted the performances of TRNSYS as a physical simulator, as well as its limits in terms of computational speed: the authors claimed that a full optimization process would take as much as ten years, had they not replaced TRNSYS with a surrogate model during optimization.
The authors of \cite{Bre2016ResidentialBD} studied the optimization of a single-family house using a combination of Energy-Plus and the NSGA-II optimization algorithm, and discussed sensitivity analysis using the Morris screening method.
Likewise, the authors of \cite{Recht2014AnalyseDL} performed sensitivity and uncertainty analysis on another building simulator known as COMFIE, and displayed its modelling performances on a passive building.

These papers demonstrate the capability of such approaches to optimize energy loads, given a building simulator.
However, because they require numerous calls to the physical simulation function, they are very computationally intensive.
Additionally, they do not leverage any data generated by a real building.

\subsection{Metamodels}
The building optimization literature has seen an increasing number of surrogate approaches, as recent sophisticated statistical models provide appealing solutions to be used in this context.
In \cite{Bre2020AnEM,Reynolds2018AZB}, statistical models were trained on a dataset sampled from EnergyPlus, allowing significant computational savings during optimization.
In \cite{Bre2020AnEM}, the authors proposed to combine NSGA-II with an artificial neural network metamodel, here a Feed Forward Network (FFN), in order to optimize the consumption of a $83\,\mathrm{m}^2$ house. Optimization was also conducted with the original building simulator, EnergyPlus, in order to compare both results and ensure that the FFN could be used as a substitute during optimization.
Similarly, \cite{Reynolds2018AZB} proposed a FFN based meta modelling approach to reduce up to 25\% the energy consumption in a small office building. EnergyPlus was used to sample a dataset for various zones of the building. 
The metamodel was tested using a 4-week long EnergyPlus simulation with variable set point temperatures and using an alternative weather file.
An example of recurrent neural architecture as a surrogate model can be found in \cite{9185769}, where the authors focused on an air-conditioning optimization problem using time series.

If these articles justify the use of metamodels, the question of which type of model to choose remains.
In an in-depth review, \cite{roman2020application} compares standard statistical models, such as polynomial regression, multivariate adaptive regression splines, Gaussian processes or Decision Trees, in the context of building performance simulation.
Artificial Neural Networks models stand out as a particularly relevant alternative, but are often presented in their most simple, time independent form, such as the FFN used in \cite{Bre2020AnEM}.
Although they may yield accurate predictions in some frameworks, these neural networks handle every time step independently, and are thus not adapted to time series problems.
They are usually substituted for their sequential counterparts, such as recurrent or convolutional based approaches, as demonstrated by the authors of \cite{SendraArranz2020ALS}. In their paper, they explored various architectures of Long Short Term Memory models, in order to predict HVAC consumption in buildings.
Therefore, designing metamodels for building calibration and optimization is likely to benefit from such recurrent and attention-based models.

Recurrent Neural Network (RNN) were first introduced as a more suited architecture for dealing with time varying input patterns \cite{Mozer1989AFB}. By replacing buffer based approaches with an updated context state, RNN are able to solve time series problems with short time dependencies, but are lackluster in problems requiring long term memory due to vanishing and exploding gradient \cite{Bengio1994LearningLD}. The Long Short Term Memory (LSTM) model proposed in \cite{Hochreiter1997LongSM} aims at bridging that gap by enforcing error flow throughout time in the network. The LSTM architecture was modified in \cite{Cho2014LearningPR} in order to simplify its implementation and improve computation times, resulting in a novel model called Gated Recurrent Unit (GRU). In parallel to these advances on recurrent architectures, Convolutional Neural Networks (CNN), rendered popular by \cite{Krizhevsky2012ImageNetCW} for image classification, have been adapted to time series problem. The approaches proposed in \cite{Jzefowicz2016ExploringTL,Kim2016CharacterAwareNL} outperformed traditional Natural Language Processing (NLP) models by replacing the embedding layer with a character-level convolutional layer. 

Recurrent and convolutional approaches coincide in that temporally close time steps data are matched together. In 2017, \cite{Vaswani2017AttentionIA} proposed an attention based approach to solve NLP tasks that consider the entire input sequence in parallel. The Transformer model is based on a self-attention mechanism, that computes an attention value for every element of a sequence with respect to all others to model their dependency. This attention mechanism allows to understand at each time step which input elements are crucial to predict the new state. This makes these networks more interpretable than their most widely-used recurrent counterparts such as LSTM or GRU networks and motivates a keen interest for such approaches to predict complex time series.

\subsection{Calibration}
Sampling inside temperatures and consumptions requires many unknown physical parameters. Instead of costly campaigns to measure these parameters, that would have to be reiterated for each new building, they may be estimated using an automatic calibration procedure by minimizing a cost function which associates, with each set of parameters, the discrepancy between the simulations and the true consumptions and temperatures, see \cite{Coakley2014ARO,Corff2018OPTIMIZINGTC}. As shown in \cite{Nagpal2019AMF}, calibration yields sufficiently accurate results for a variety of different buildings. This ensures limited additional costs to fit a trained metamodel to new buildings. In many related works, this problem cannot be solved since no real data are used in the calibration step, see \cite{Bre2020AnEM,Reynolds2018AZB}, i.e. the calibration is performed based only on simulated data. Optimization is thus conducted without justifying that the theoretical energy savings could be applied to any real building.

The calibration task revolves around a non differentiable optimization problem, which is often tackled by using genetic optimization methods. In \cite{Aird2016APPLICATIONOA}, the authors demonstrate the use of the Non-dominated Sorting Genetic Algorithm II (NSGA-II) to select a set of estimated parameters that jointly minimize the coefficient of variation of the root mean square error, and the normalized mean bias error. 
All criteria can instead be combined in a single calibration error, in order to turn to single objective differentiation free algorithms that offer a single best candidate, avoiding the need for further selection processes. In \cite{Corff2018OPTIMIZINGTC}, the CMA-ES algorithm introduced in \cite{igel:hansen:roth:2007} was used to minimize a combination of heating and cooling errors.

\subsection{Our approach}
In this paper, we propose an end-to-end methodology, from dataset sampling to metamodel calibration and optimization using data obtained from wireless sensors set in large buildings.
We introduce a new metamodel to predict building behaviors after a comprehensive study of several approaches from traditional RNN to a model based on a Transformer architecture \cite{Vaswani2017AttentionIA}.
The performance of this metamodel are compared both in terms of accuracy and computational efficiency with TRNSYS.

Once the metamodel is trained using a dataset built using TRNSYS, all the parameters of real buildings and of their Building Management System (BMS) are estimated using real measurements with the CMA-ES algorithm.
A multi-objective methodology to improve energy efficiency and maintain thermal comfort is then implemented by acting only on the BMS.
The NSGA-II approach is used to obtain a Pareto front, i.e. a set of optimal compromises between consumption and comfort, whereby improving one criteria leads to necessarily degrading the other.
Our methodology is summarized in a flowchart in the appendix, see Figure~\ref{fig:flowchart}.

The paper is organized as follows: Section~\ref{sec:models} provides all the deep learning architectures used in this paper to build a metamodel.
It also describes the data and variables used in our metamodel.
Section~\ref{sec:calib:optim} illustrates the performance of our metamodel in the calibration and optimization processes for two real buildings.
The numerical experiments illustrate how the same metamodel ensures a significant gain in energy in various settings in comparison to the considered alternatives.

\section{Meta modelling}
\label{sec:models}

\subsection{Notations}
Let $(Y_k)_{k\geq 0}$ be a multivariate variable describing the state of the building at each hour, denoted by the index $k$.
Based on the modeling of an equivalent building by energy managers, this variable contains $8$ time series, such as inside temperatures, heating, cooling and ventilation consumptions, see Table~\ref{tab:Yk} for a detailed list.
The aim of the metamodel introduced in this paper is to provide a numerically efficient solution to predict $(Y_k)_{k\geq 0}$ from several sets of input variables.

Input time series are divided in two variables: $(\varphi_k, \psi_k)_{k\geq 0}$.
The input $\varphi_k$ contains weather data at time $k$, such as outdoor temperature, relative humidity or irradiance values, see Table~\ref{tab:phi}; $\psi_k$ stores the information relative to the building usage: activation hours of the AC and ventilation systems, comfort and reduced temperatures (thermal objectives when the building is occupied or empty), see Table~\ref{tab:psi}. The parameter $\lambda$ represents all unknown parameters regarding the geometrical description of the buildings (windows area ratio, etc.), as well as parameters related to heat transfer (heat capacity, infiltration rate, etc.), and occupation schedules, see Table~\ref{tab:lambda}.
We assume in this study that this parameter set does not evolve with time.

In this section, we describe how a simulator may be used to train the metamodel which aims at mimicking its outputs for various choices of $\lambda$, $(\psi_k)_{k\geq 0}$, and of meteorological data $(\varphi_k)_{k\geq 0}$.

\subsection{Proposed benchmarks}
In most recent works, a great deal of research activities focused on FFN as surrogate models, see \cite{Bre2020AnEM,Magnier2010MultiobjectiveOO,Reynolds2018AZB}.
Although they may lead to interesting performance during the training phase, these fully connected architectures are not well suited for time series prediction, in particular for long time spans.
We ceased this opportunity to explore other approaches that have proven to be more relevant for solving time series tasks in the past few years.
Therefore, we decided to evaluate the go-to architectures for time series: a bidirectional LSTM, a bidirectional GRU (BiGRU), a hybrid model mixing both convolutional and GRU layers (ConvGru) and a Feed Forward Network (FFN) as used in previous works.
In addition to those models, a Transformer model, which introduces an attention mechanism to model dependencies, is also considered.
These models have been implemented using the deep learning framework PyTorch.

\subsection{Our proposed metamodel}
\label{sec:metamodel}
Our metamodel is a function $f_{\theta}: (h_{k-1},u_k) \mapsto y_k$ with parameters $\theta$ that maps, at each time step $k$, the building state $u_k \equiv (\varphi_k, \psi_k, \lambda)$ and a hidden state $h_{k-1}$ depending on the past values $(u_1,\ldots,u_{k-1})$, to a prediction of its indoor temperature and consumptions $y_k$.
The model is trained to produce accurate predictions by tuning its parameters $\theta$, usually referred to as weights, through an iterative back propagation algorithm, where predictions $y_k$ are compared to the ground truth $Y_k$.

We use as a backbone a many to many RNN architecture, and denote by $h_k^\ell$ and $x_k^\ell$ the hidden state and input of layer $1 \leq \ell \leq L$ at time step $k$, in particular $x_k^0 \equiv u_k$.
The hidden state is traditionally initialized as the zero vector, $h_0^\ell \equiv 0$ for all $1 \leq \ell \leq L$.

\begin{figure}[htpb]
	\centering
	\includegraphics[width=0.45\textwidth]{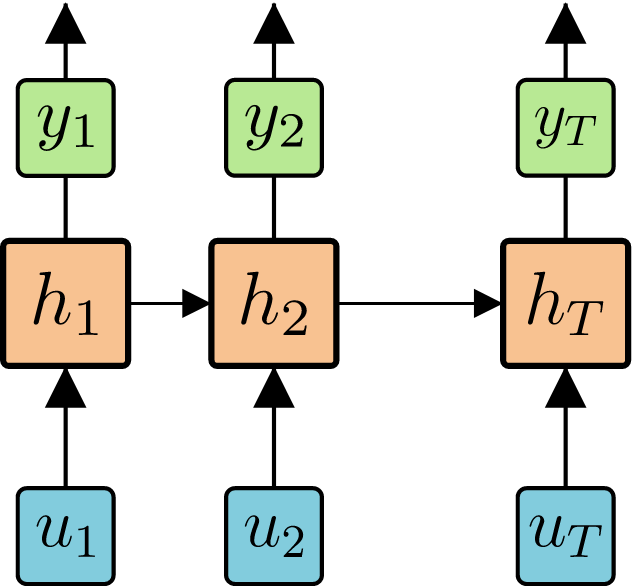}
	\includegraphics[width=0.45\textwidth]{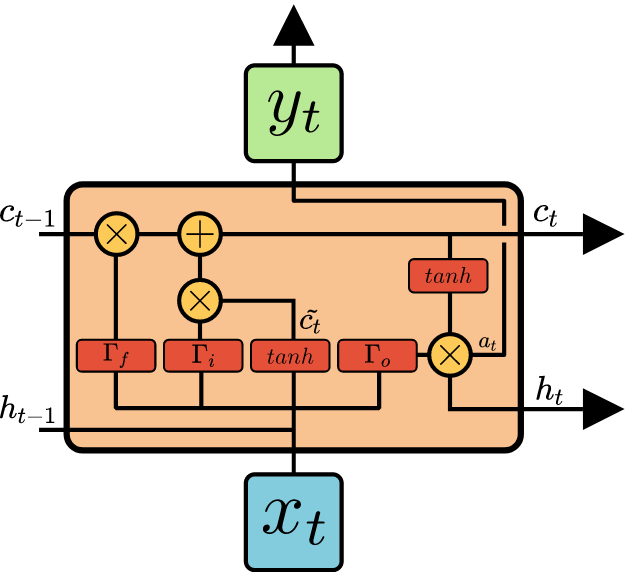}
	\caption{Our metamodel architecture (left), and a detailed LSTM cell (right). The LSTM cell improves on the classic RNN by introducing a cell state $c_t$ supposed to carry long term memory, without additional alterations, throughout the sequence. The three input gate $\Gamma_i$, forward gate $\Gamma_f$ and output gate $\Gamma_o$ determine whether information in both hidden state $h_k$ and cell state $c_k$ should be carried away or discarded.}
	\label{fig:architecture}
\end{figure}

In the original and most simple definition of a RNN, the hidden state is computed recursively as $h_k^\ell = \tanh(W_{ih}^\ell x_k^\ell + W_{hh}^\ell h_{k-1}^\ell + b_h^\ell)$, where $W_{ih}$, $W_{hh}$ and $b_h$ are the weight matrices and bias learned during training, and initialized with random values. Our metamodel is based on a LSTM architecture and replaces the update of the hidden state by the following state equations:
\begin{align*}
	\Gamma_i^\ell & = \sigma(W_{xi}^\ell x_k^\ell + W_{hi}^\ell h_{k-1}^\ell + b_i^\ell)\,, \\
	\Gamma_f^\ell & = \sigma(W_{xf}^\ell x_k^\ell + W_{hf}^\ell h_{k-1}^\ell + b_f^\ell)\,, \\
	\Gamma_o^\ell & = \sigma(W_{xo}^\ell x_k^\ell + W_{ho}^\ell h_{k-1}^\ell + b_o^\ell)\,, \\
	\tilde c_k    & = \tanh(W_{xc}^\ell x_k^\ell + W_{hc}^\ell h_{k-1}^\ell + b_c^\ell)\,,  \\
	c_k^\ell      & = \Gamma_f^\ell * c_{k-1}^\ell + \Gamma_i^\ell * \tilde c_k\,.          \\
	h_k^\ell      & = \Gamma_o^\ell * \tanh c_k^\ell\,.
\end{align*}
An additional fully connected layer is added on top of the RNN architecture, following results presented in \cite{SendraArranz2020ALS}:
$$y_k = \sigma (W_y h_k^L + b_y)\,,$$
where $\sigma$ is the sigmoid activation function $\sigma: x \mapsto (1 + e^{-x})^{-1}$.
The architecture is represented in Figure~\ref{fig:architecture}. The parameters to be estimated during the training phase of the metamodel are
$$
	\theta = \left\{\left(W_{xi}^\ell,W_{hi}^\ell,W_{xf}^\ell,W_{hf}^\ell,W_{xo}^\ell,W_{ho}^\ell,W_{xc}^\ell,W_{hc}^\ell,W_y,b_{i}^\ell,b_{f}^\ell,b_{o}^\ell,b_{c}^\ell,b_y\right)_{1\leq \ell \leq L}\right\}\,.
$$

\subsection{Dataset sampling}
\label{sec:dataset}
The training dataset is sampled by exploring the input space of the simulator.
We chose TRNSYS in this paper but any simulator can be used to train the metamodel.

We define ranges for each input variable in $\lambda$ and $(\psi_k)_{k\geq 0}$ with the help of energy managers, such as highest and lowest scheduled temperature, or the most early and late hour of arrival of occupants, see the appendices for a complete list of these ranges.
Because our dataset aims at capturing multiple buildings, these ranges are not centered around a specific set of variables, but rather cover all possible values across our cluster of buildings.
In addition, real weather data $(\varphi_k)_{k\geq 0}$ acquired between June and August 2020 around the Parisian area where used to obtain a dataset consistent with the real buildings.

In our numerical experiments, we chose a uniform sampling method over the ranges for each building and weather variable.
This allows us to easily split the dataset uniformly into training and testing sets, which is crucial to validate the metamodel.

The input vector $u_k$ contains 27 variables at each time step: 8 variables from $\lambda$, 7 from $\varphi_k$ and 12 from $\psi_k$.
A total of 15,000 training examples were sampled, an example being a month i.e. 672 hours.
During the training phase, the parameters of each metamodel described in Section~\ref{sec:models}, and called $\theta$ in the detailed case of the RNN approach, are estimated based on this dataset.
The metamodels compared in this section are defined with a latent dimension of $d_{emb} = 64$ and a total of $L=4$ layers.
Hyper parameters, such as learning rate, dropout, number of epochs or batch size, were chosen empirically by interpreting the cost history of the model during the training, and the learning curve.


\subsection{Training}
\label{sec:training}
During training, for each example, we compute the Mean Squared Error (MSE) loss, and combine consumption and temperature errors:
\begin{align*}
	\mathrm{MSE_T} = \frac{\sum_{k=1}^M (T_k - \widehat T_k)^2}{\sum_{k=1}^M (T_k - \overline T)^2}\quad\mathrm{and} & \quad
	\mathrm{MSE_Q} = \frac{\sum_{k=1}^M (Q_k - \widehat Q_k)^2}{\sum_{k=1}^M (Q_k - \overline Q)^2}                                                    \\
	\overline T = \frac{1}{M}\sum_{k=1}^M T_k \quad \mathrm{and}                                                     & \quad
	\overline Q = \frac{1}{M}\sum_{k=1}^M Q_k                                                                                                          \\
	\mathrm{loss} = \beta \mathrm{MSE_T}                                                                             & + (1 - \beta) \mathrm{MSE_Q}\,,
\end{align*}
where $M$ is the number of data in each example, $T_k$ and $Q_k$ are the ground truth at time $k$, and $\widehat T_k$ and $\widehat Q_k$ are the predictions given by the metamodel with the current value $\theta$ of the metamodel for temperature and total consumption respectively.
In the experiments below, as the inside temperatures and all consumptions are normalized, we chose the non informative value $\beta=0.5$.
We chose the Adam optimizer \cite{Kingma2015AdamAM} and all simulations were computed on a single 1080TI GPU card.

\subsection{Validation}
Validation is essential to identify any potential overfit of the model on the training dataset.
In this study, we implement a traditional cross-validation, whereby the dataset is split into $k$ folds, and the model is trained on the $k-1$ first folds and evaluated on the last.
We average this validation score by iteratively changing the validation fold, as detailed by the authors of \cite{Seyedzadeh2020MachineLM}, with $k=5$.
This method ensures that our model is always evaluated on unseen data, which demonstrates is generalization capability and avoids any potential bias of the validation split.
Table~\ref{table:train} displays the mean values and standard deviations of the loss function of this cross validation at the end of the training procedure.
The table also displays the mean squared error $\mathrm{MSE_T}$ (resp. $\mathrm{MSE_Q}$) on the temperatures (resp. consumptions) only, as well as these same metrics computed only during occupation time: $\mathrm{MSE_T^{occ}}$ and $\mathrm{MSE_Q^{occ}}$.
For a global consumption evaluation, we compute the absolute relative error on the cumulative consumptions $\Delta_{Q^{\mathrm{Tot}}}$.

\begin{table}
	\caption{Metrics (means and standard deviations) of the metamodels on the validation splits. The best mean values are displayed in bold (the lowest losses and mean squared errors). $\mathrm{Time}$ is the computation time to run a single simulation through the network, and was estimated by averaging 100 inferences. Our selected architecture is detailed in Section~\ref{sec:metamodel}, and achieves the best performances on all metrics, while still being on par with most models in computation time.}
	\label{table:train}
	\centering
	\resizebox{\textwidth}{!}{
		\begin{tabular}{ cccccc }
			\hline
									& BiGRU              & Transformer        & Ours                         & ConvGru            & FFN                \\
									\hline
			$\mathrm{Loss}$ \;\;\;\;\;($\times 10^{-4}$)    & $2.05\;  (1.61)$ & $2.72 \; (2.88)$ & $\textbf{1.65  (1.47)}$  & $2.26 \; (2.04)$ & $90.5 \; (41.1)$ \\
			$\mathrm{MSE_T}$ \;\;($\times 10^{-5}$)         & $1.00\;  (1.53)$ & $1.49 \; (2.75)$ & $\textbf{0.820  (1.44)}$ & $1.20 \; (2.05)$ & $39.5 \; (39.7)$ \\
			$\mathrm{MSE_Q}$ \;\;($\times 10^{-4}$)         & $3.84\;  (2.05)$ & $4.12 \; (3.01)$ & $\textbf{2.75  (1.72)}$  & $3.53 \; (1.77)$ & $172  \;(60.2)$  \\
			$\mathrm{MSE_T^{occ}}$ ($\times 10^{-5}$)       & $4.60\;  (7.16)$ & $6.56 \; (12.2)$ & $\textbf{3.79  (6.81)}$  & $5.89 \; (9.91)$ & $176  \;(189)$   \\
			$\mathrm{MSE_Q^{occ}}$  ($\times 10^{-4}$)      & $1.45\;  (1.00)$ & $2.07 \; (1.80)$ & $\textbf{1.22  (0.878)}$ & $1.85 \; (1.23)$ & $113  \;(50.8)$  \\
			$\Delta_{Q^{\mathrm{Tot}}}$  ($\times 10^{-3}$) & $4.03\;  (11.7)$ & $20.1  (12.4)$ & $\textbf{2.46 (12.0)}$  & $14.9 \; (16.2)$ & $1.82 \; (68.8)$ \\
			$\mathrm{Time}\;\;\;\;\; (\times 10^{-2}s)$     & 6.46               & 4.52               & 6.51                         & 6.77               & \textbf{0.341}     \\
			\hline
		\end{tabular}}
\end{table}

\section{Energy Optimization in real buildings}
\label{sec:calib:optim}
The experiments conducted in our paper to analyze the performance of the trained metamodel focused on the optimization of two buildings in the Parisian region.
Each one is represented by a single thermal zone.

\begin{itemize}
	\item Stanley is a $18,512\, m^2$ building. It is delimited by four vertical walls of dimension $2,314\, m^2$, $1,917\, m^2$, $2,123\, m^2$ and $1,725\, m^2$, as well as a roof and ground of dimension $2,304\, m^2$. The main insulator is a $10\, cm$ layer of polystyrene. It was built in 1983.
	\item Livingstone is a $13,594\, m^2$ building, including 4 vertical walls with respective areas $1,678\, m^2$, $1,274\, m^2$, $1,281\, m^2$ and $1,252\, m^2$, a horizontal roof and a horizontal ground of dimension $4,653\, m^2$ and $4,286\, m^2$. The main insulator is a $8\, cm$ layer of polyurethane. It was built in 2006.
\end{itemize}

Based on a commonly used rule, it is assumed that $2/3$ of the full area is occupied by people. Assuming that each occupant requires $12\, m^2$, this allows to set the initial values for the number of occupants and the number of PCs (set to 1.2 times this value) in the building during occupancy hours. These values are assumed to be known and fixed and used to sample the training dataset.

\subsection{Calibration}
During the training phase, metamodel parameters are estimated by minimizing the loss function on the simulated dataset which corresponds to various configurations associated with choices of $\lambda$ and $(\psi_k, \varphi_k)_{k\geq 0}$.
Because this dataset is sampled from a simulation model, we trained the metamodel ignoring real building related noise and measurement errors.
Additionally, both the BEM and our metamodel require $\lambda$ that cannot be properly identified for each building especially without renovation work.
By comparing the metamodel predictions to real historic data during the calibration phase, we search for a set of building related parameters that best match reality.
During this step, the weights $\theta$ of the metamodel are frozen, meaning that we no longer update each weight matrix of the neural network.

We can compute, for each given set of input parameters $\lambda$ and $(\psi_k, \varphi_k)_{k\geq 0}$, the difference between estimated and real historical data.
Goodness of fit of the model is measured with a normalized Mean Square Error denoted $\Delta_{\texttt{calib}}$ defined in \eqref{eq:delta_calib}, following the performance evaluation criteria in \cite{Ajib2018DatadrivenBT}.
Because this is a non differentiable problem, the cost function cannot be minimized using a stochastic gradient descent algorithm as in the training step; instead we use the CMA-ES algorithm \cite{Hansen2016TheCE}, an evolutionary algorithm designed to solve constrained non-convex optimisation problems.
In our experiments, the variables we adjust for fitting are constrained by the same ranges defined in the data sampling section.
The algorithm is implemented by the author of the paper in the pycma library\footnote{https://github.com/CMA-ES/pycma}.

Following traditional methodology in building calibration, we measure the performances of the calibrated model with the Mean Bias Error ($\mathrm{MBE}$) and Coefficient of variation of the Root Mean Square Error ($Cv(\mathrm{RMSE})$) criteria.
For any sequence $(z_k)_{1\leq k\leq M}$ associated with predictions $(\widehat{z}_k)_{1\leq k\leq M}$, these quantities are defined as follows:
\begin{align}
	\mathrm{MBE} (\%) & = 100\frac{\sum_{k=1}^M{\left(z_k - \widehat{z}_k\right)}}{\sum_{k=1}^M{z_k}}\,, \quad
	Cv(\mathrm{RMSE}) (\%) = 100\frac{\mathrm{RMSE}}{\overline{z}}\,, \label{eq:mbe}                           \\
	\mathrm{RMSE}     & = \left( \frac{\sum_{k=1}^M{(z_k-\widehat{z}_k)^2}}{M}\right)^{1/2} \,, \quad
	\overline{z} = \frac{1}{M}\sum_{k=1}^M z_k \nonumber
\end{align}
\begin{equation}
	\Delta_{\texttt{calib}} = \frac{\sum_{k=1}^M (z_k - \widehat{z}_k)^2}{\sum_{k=1}^M (z_k - \overline{z})^2}
	\label{eq:delta_calib}
\end{equation}
where $M$ is the number of data in each example.
In a detailed review of calibration methods, the authors of \cite{Fabrizio2015MethodologiesAA} have gathered the international recommended ranges regarding these criteria, when validating a calibrated model.
Regardless of the simulation program, the $Cv(\mathrm{RMSE})$ should fall within $\pm 20\%$, and the $\mathrm{MBE} \pm 5\%$ when considering hourly calibrations.
As shown in Table~\ref{tab:calib}, our results for both consumptions and indoor temperatures calibration are well within these guidelines.

\begin{figure}[htpb]
	\centering
	\includegraphics[width=\textwidth]{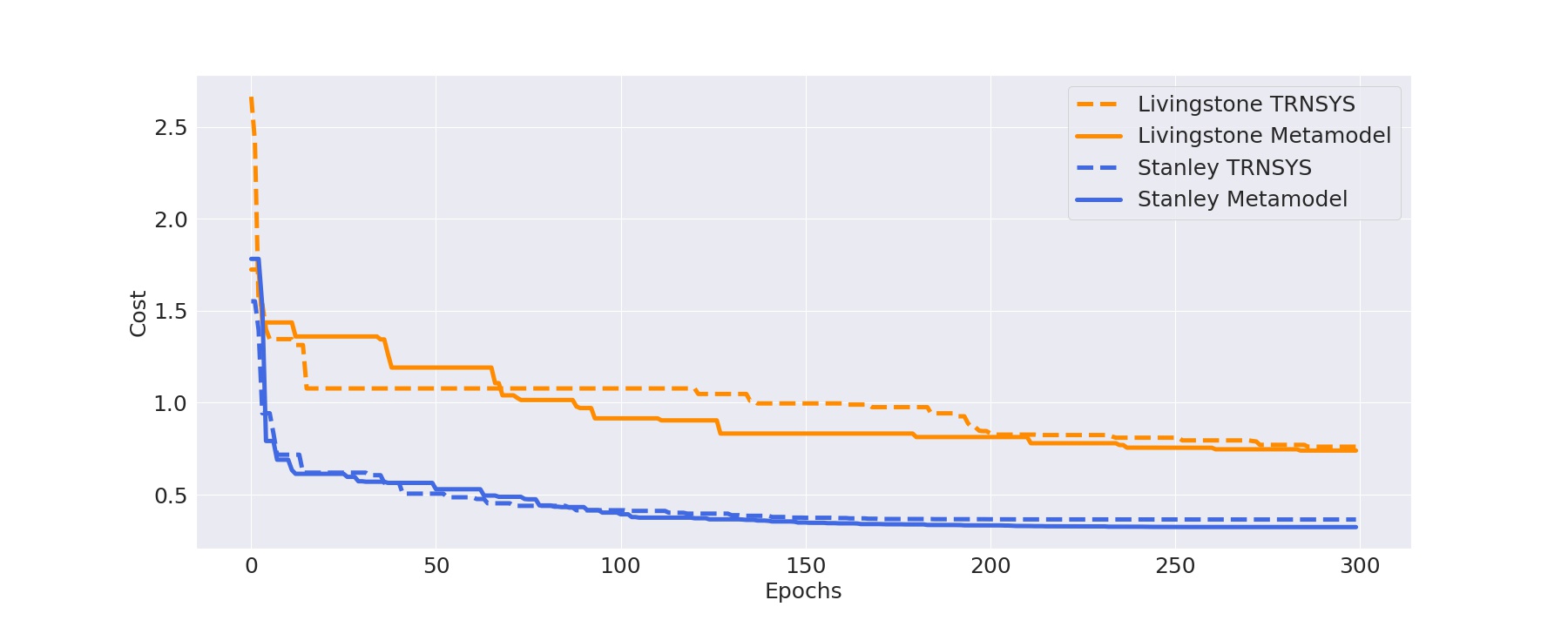}
	\caption{Calibration cost evolution for the metamodel and TRNSYS ($\Delta_{\texttt{calib}}$, see \ref{eq:delta_calib}), on Livingstone and Stanley. Both models were calibrated for 300 epochs, which is enough to reach convergence for Stanley, but not Livingstone.}
	\label{fig:calib_convergence}
\end{figure}

\begin{figure}[htpb]
	\centering
	\includegraphics[width=0.95\textwidth]{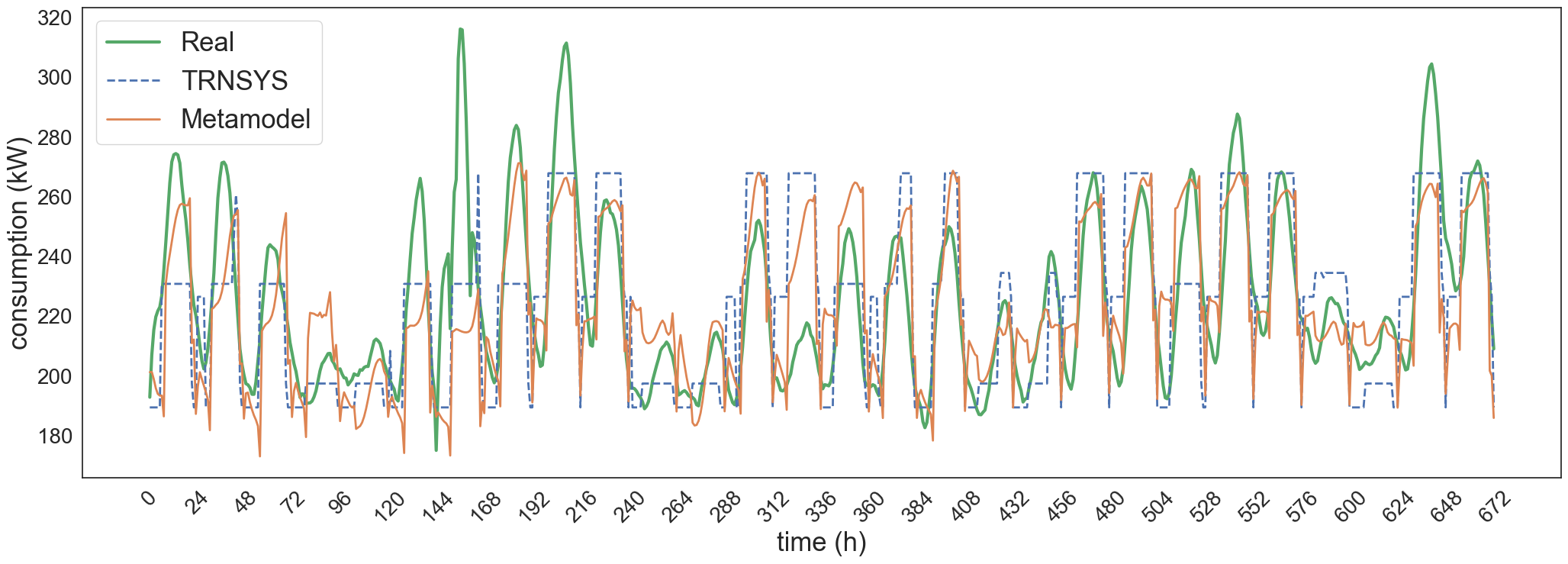}
	\includegraphics[width=0.95\textwidth]{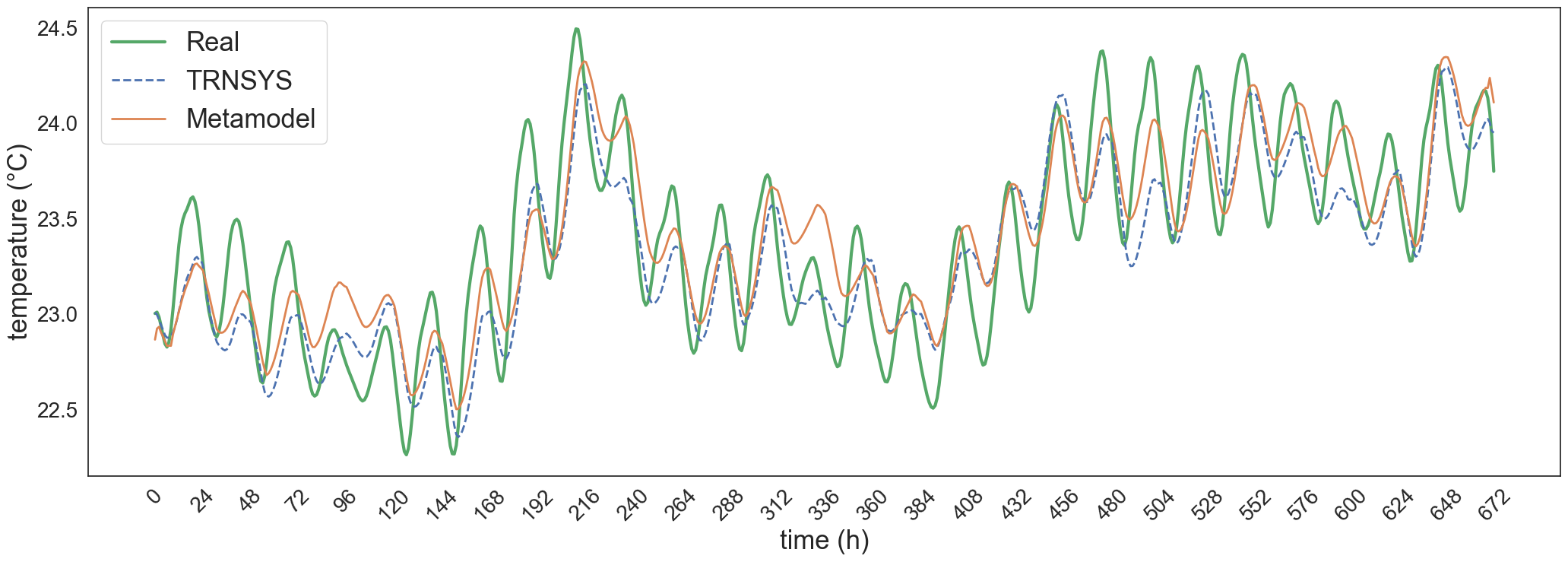}
	\caption{Consumption and temperature simulations after calibration, for both the metamodel and TRNSYS, for Stanley.}
	\label{fig:calib_comparison_stanley}
\end{figure}

\begin{figure}[htpb]
	\centering
	\includegraphics[width=0.9\textwidth]{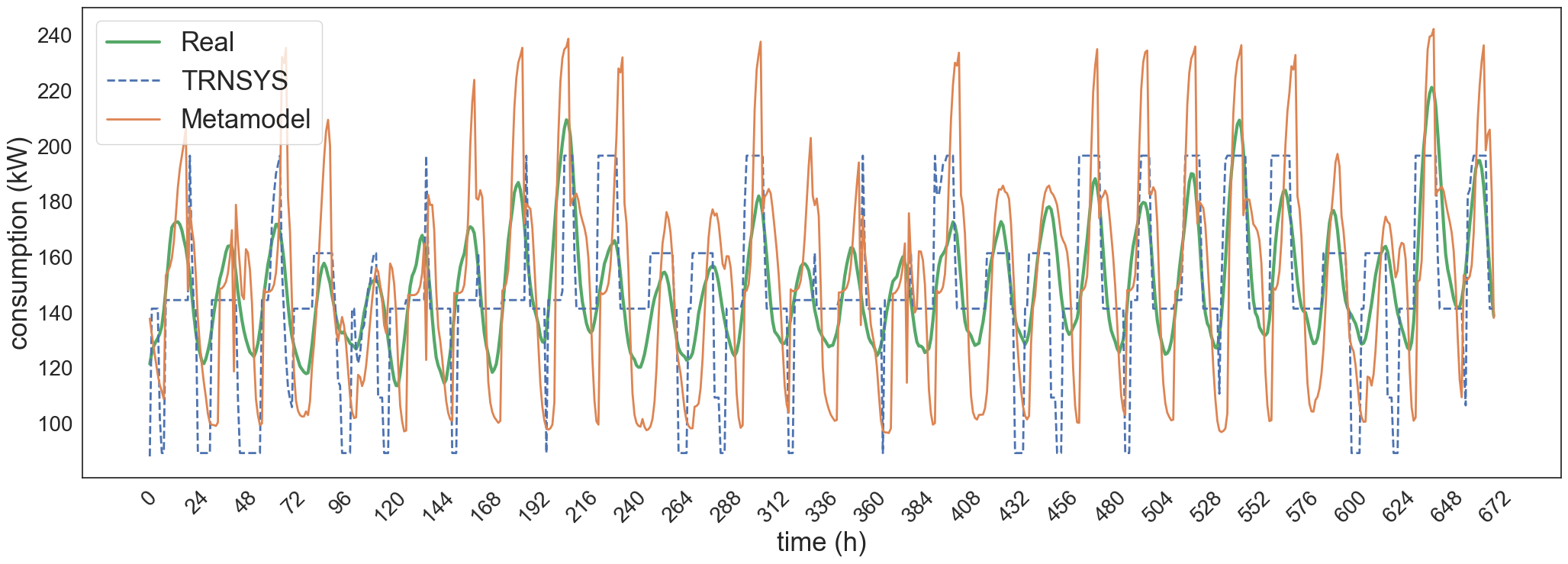}
	\includegraphics[width=0.9\textwidth]{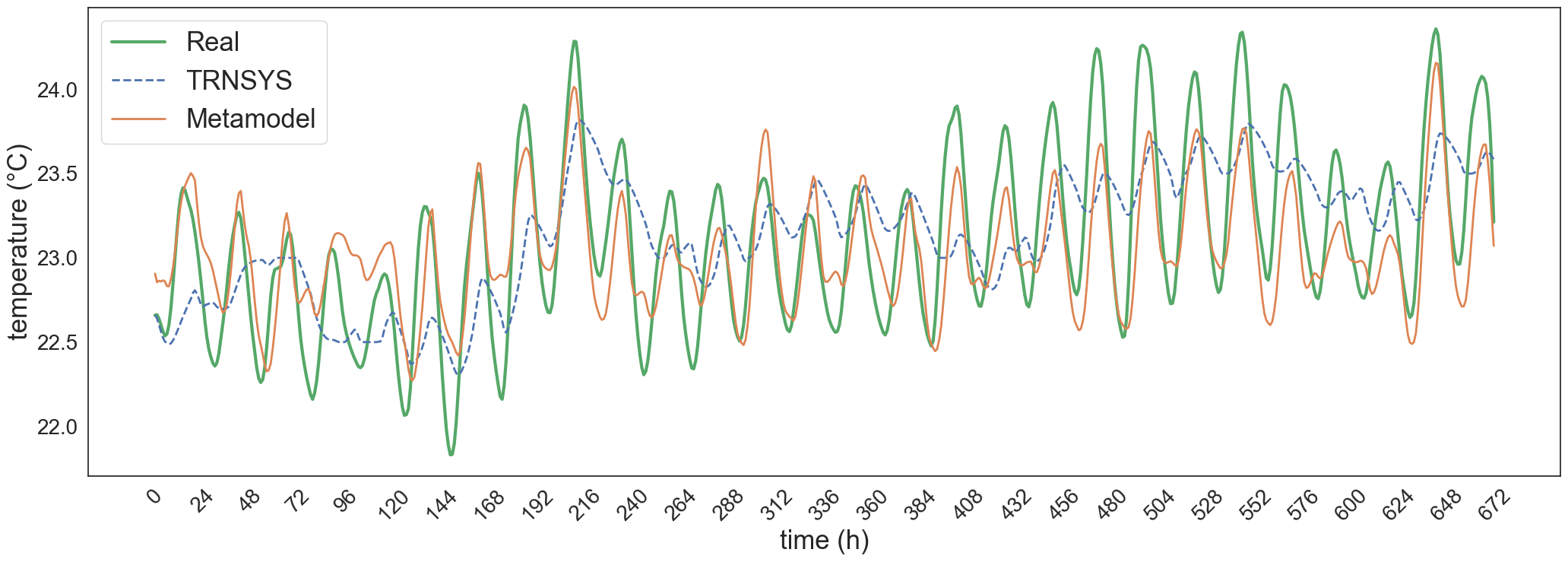}
	\caption{Consumption and temperature simulations after calibration, for both the metamodel and TRNSYS, for Livingstone.}
	\label{fig:calib_comparison_livingstone}
\end{figure}

Calibration was run for both the metamodel and the original BEM (TRNSYS) for a maximum of 3 hours.
As shown in Table~\ref{tab:calib}, we can achieve satisfactory results for Stanley in this timespan, as both model converge to close values for both the $Cv(\mathrm{RMSE})$ and $\mathrm{MBE}$.
Figure ~\ref{fig:calib_comparison_stanley} displays both models calibration results, compared to real data.
On the other hand, TRNSYS calibration of Livingstone is sensibly below the results obtained with the metamodel, as calibration did not converge in the available time, see Figure~\ref{fig:calib_convergence}.
The calibration of the metamodel reached convergence but with a tremendous number of epochs, that would have required to run TRNSYS for about 10 hours in order to get similar performances.
As a comparison, we calibrated the metamodel for the same number of epochs as TRNSYS, and obtained similar results.
This experiment comforts the idea that TRNSYS and the metamodel behave similarly after the calibration step, but the much shorter computation time of the metamodel allows us to better calibrate complex buildings, such as Livingstone.
See Figure ~\ref{fig:calib_comparison_livingstone} for a visualization of the TRNSYS and metamodel calibration after one hour.

\begin{table}[htpb]
	\centering
	\caption{Calibration metrics for Stanley and Livingstone buildings, see \ref{eq:mbe}.
		Convergence is reached for Stanley after $300$ iterations, which is not enough for Livingstone, as displayed in Figure~\ref{fig:calib_convergence}.
		This table demonstrates that the metamodel and TRNSYS perform similarly when calibrated for the same number of iterations, although the metamodel is much faster.
		Additionally, only the metamodel is able to reach convergence for Livingstone in a reasonable time frame.}

	\label{tab:calib}
	\resizebox{\textwidth}{!}{
		\begin{tabular}{*7c}
					  & $\mathrm{MBE}_Q$ & $Cv(\mathrm{RMSE})_Q$ & $\mathrm{MBE}_T$ & $Cv(\mathrm{RMSE})_T$ & Iterations & Computational time \\
			\hline
			{\bf Stanley}     &                  &                       &                  &                       &            &                    \\
			Metamodel         & -0.627           & {\bf 11.0}            & 0.134            & {\bf 1.20}            & 300        & 2mn                \\
			TRNSYS            & -0.409           & 12.1                  & -0.264           & 1.24                  & 300        & 3h                 \\
			\hline
			{\bf Livingstone} &                  &                       &                  &                       &            &                    \\
			Metamodel         & -0.690           & {\bf 14.2}            & -0.0551          & {\bf 1.29}            & 10000      & 1h                 \\
			Metamodel         & -0.574           & {\bf 14.2}            & -0.413           & 1.95                  & 300        & 2mn                \\
			TRNSYS            & -1.08            & 15.8                  & 0.156            & 1.96                  & 300        & 3h                 \\
		\end{tabular}
	}
\end{table}

\paragraph{Validation}
\label{calib:validation}
The metamodel will assist the decision process for building management by simulating thermal behavior of future weeks.
Because the calibration process requires real data, the metamodel is calibrated on several past weeks, in order to capture the real building behavior in a situation as close as possible to the future period we aim to match.

\begin{figure}[htpb]
	\centering
	\includegraphics[width=0.85\textwidth]{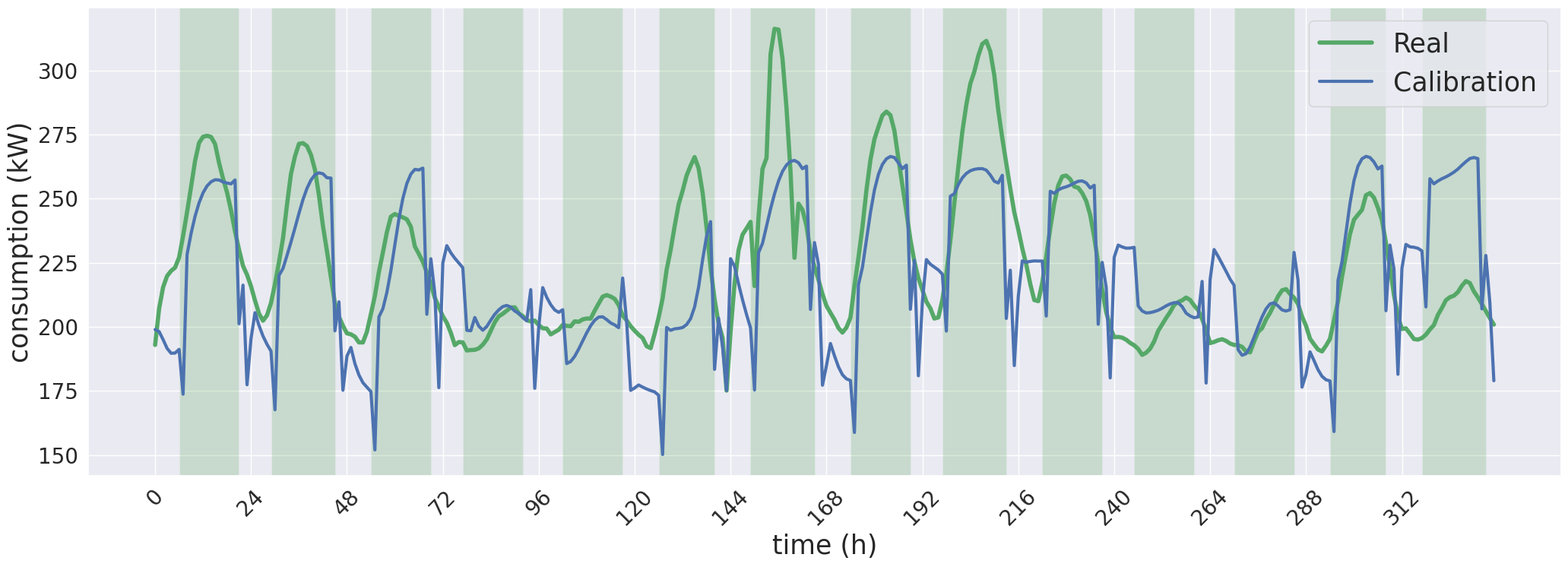}
	\includegraphics[width=0.85\textwidth]{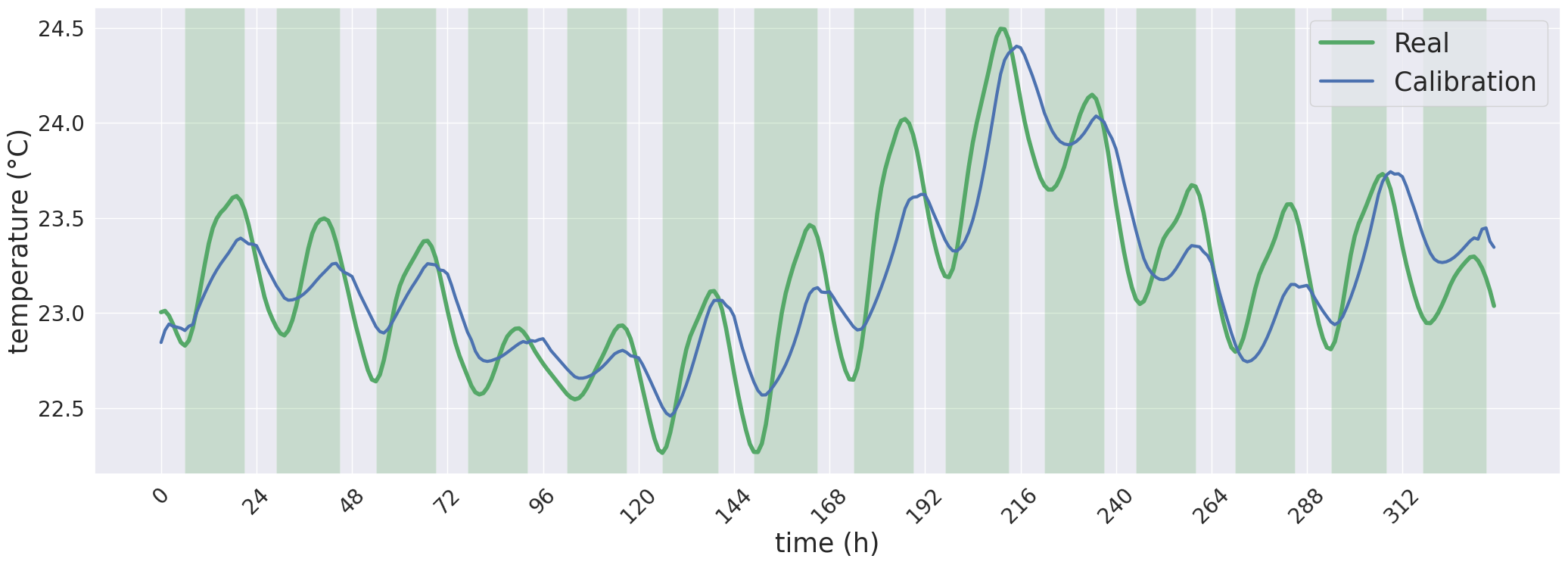}
	\includegraphics[width=0.85\textwidth]{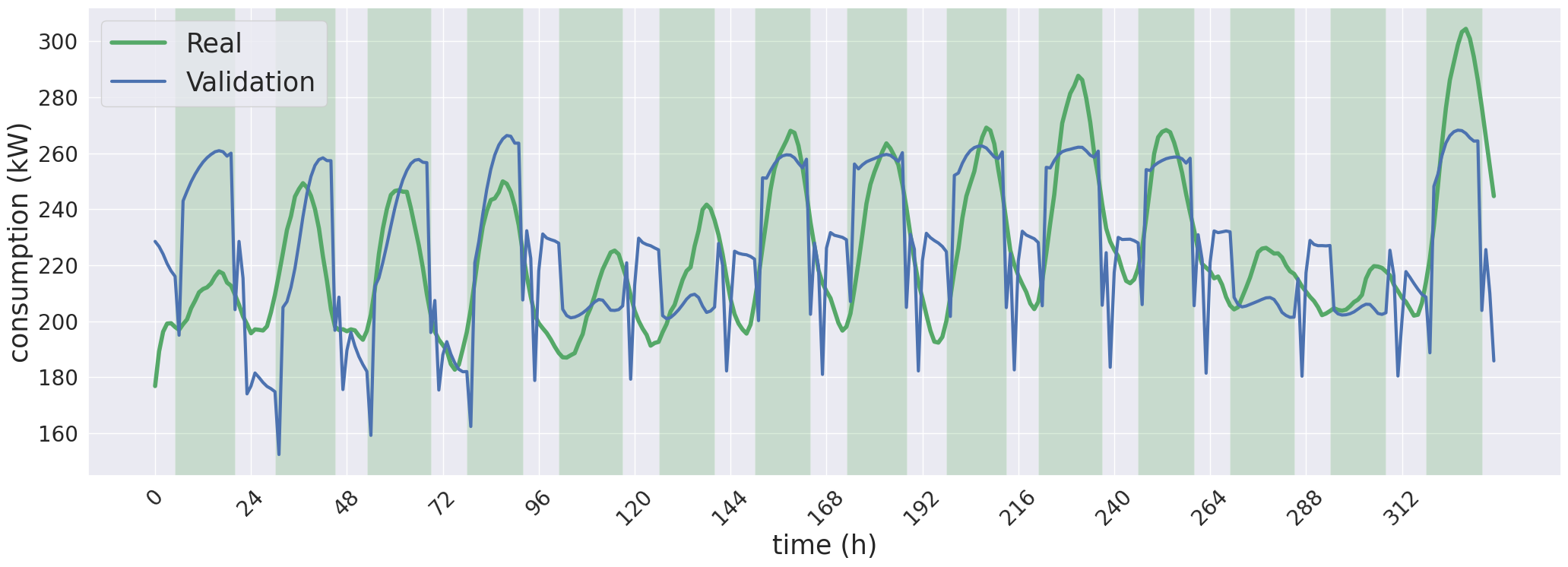}
	\includegraphics[width=0.85\textwidth]{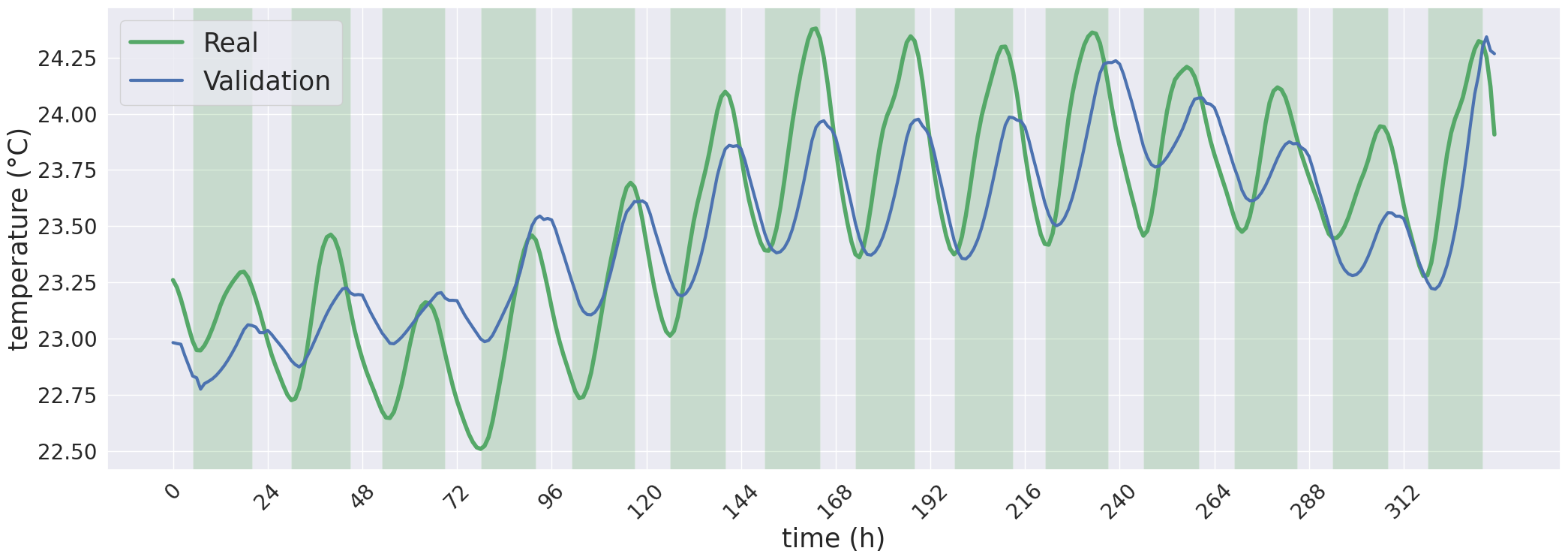}
	\caption{Consumption and temperature simulations after calibration on two weeks (top), simulation on the two following weeks for the same parameters (bottom), for Stanley. Although curves for the validation weeks are not matched perfectly, the metamodel is able to capture most trends of both consumption and indoor temperature. The remaining difference can be explained by the absence of real building usage settings $\psi$ for the calibration week. This experiment comforts our assumption that the calibration step leads to a correct estimation of building parameters $\lambda$.}
	\label{fig:calib_validation}
\end{figure}

We validate the calibration phase using two successive weeks, by applying the calibrated settings to the two following weeks, with fresh weather data, and compare the results to the true observed values.
The results are displayed in Figure~\ref{fig:calib_validation} and display encouraging results, as the simulation of the metamodel on the two unseen weeks is able to match most trends is both consumption and indoor temperature.

\subsection{Optimization}
After a successful calibration, the metamodel is supposed to have correctly estimated building parameters $\lambda$, enabling it to accurately reproduce the thermal exchanges of the real building, as confirmed by the validation step.

\begin{figure}[htpb]
	\centering
	\includegraphics[width=0.45\textwidth]{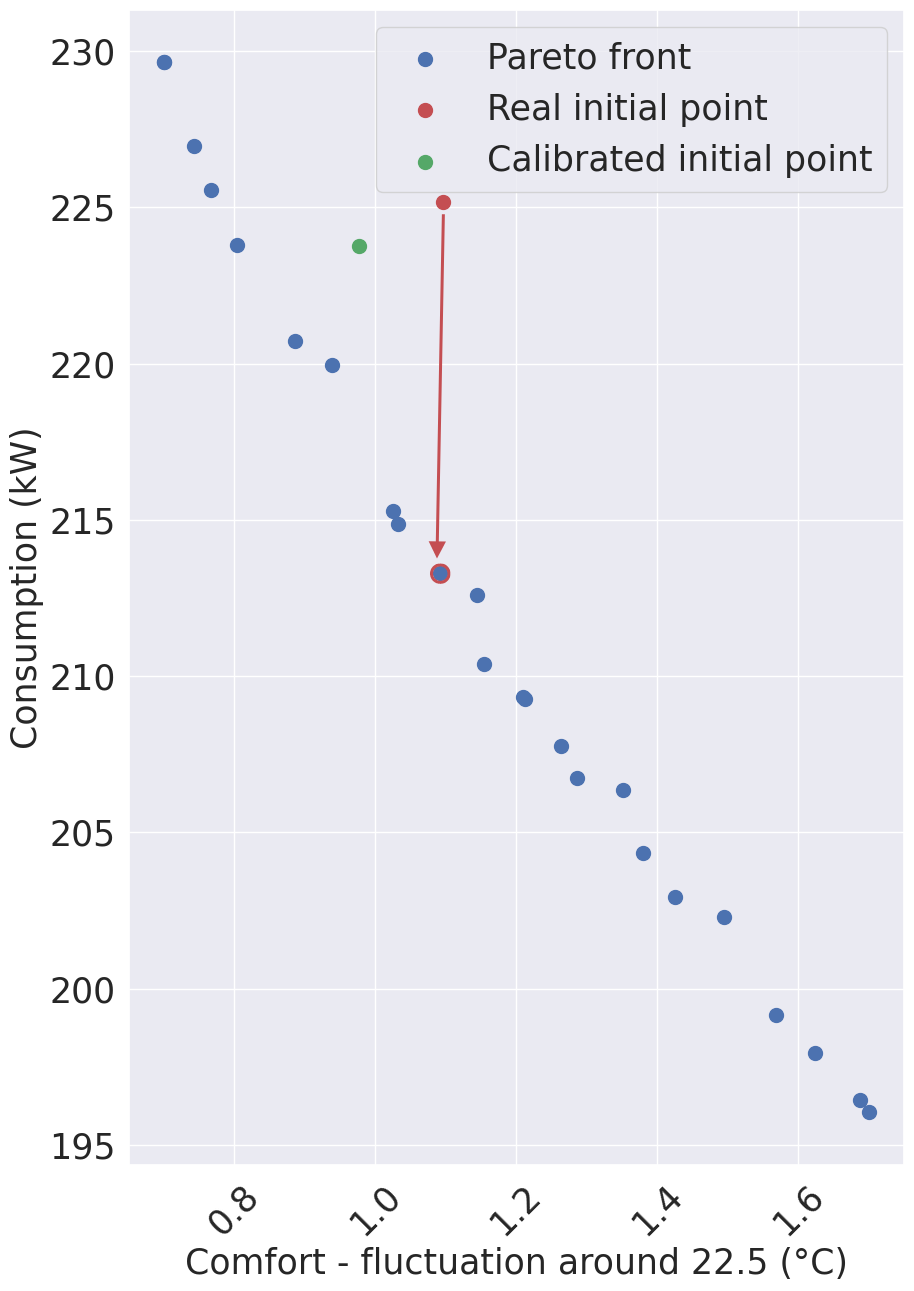}
	\includegraphics[width=0.45\textwidth]{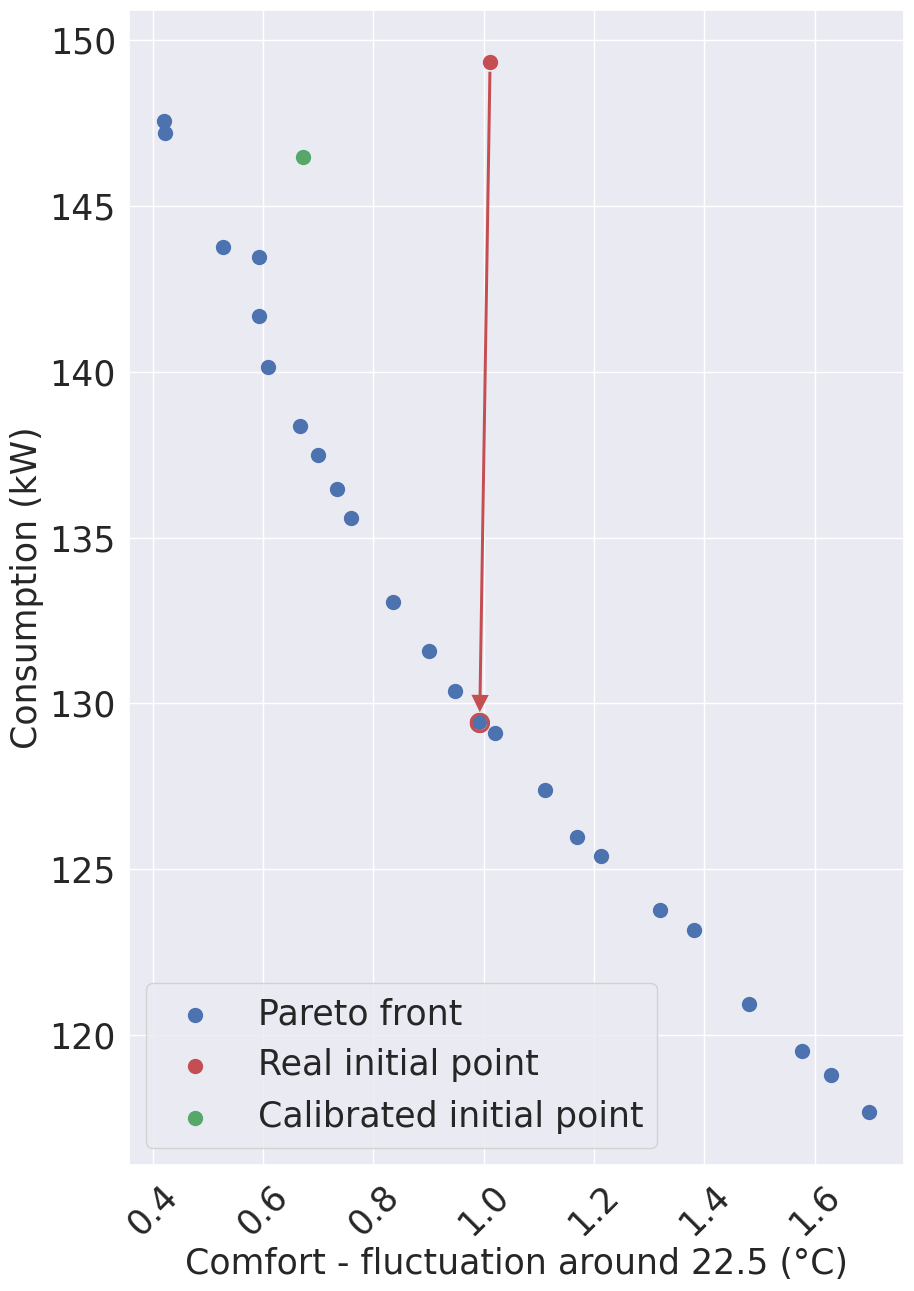}
	\caption{Pareto front after optimization for the Stanley (left) and Livingstone (right) building. We select the point of closest equivalent comfort, corresponding to a 5.3\% (Stanley) and 9.9\% (Livingstone) reduction in consumption. Combinations of energy consumption and comfort are unreachable below the Pareto front, and suboptimal above; we always aim at sampling points at the intersection.}
	\label{fig:pareto}
\end{figure}

\begin{figure}[htpb]
	\centering
	\includegraphics[width=0.95\textwidth]{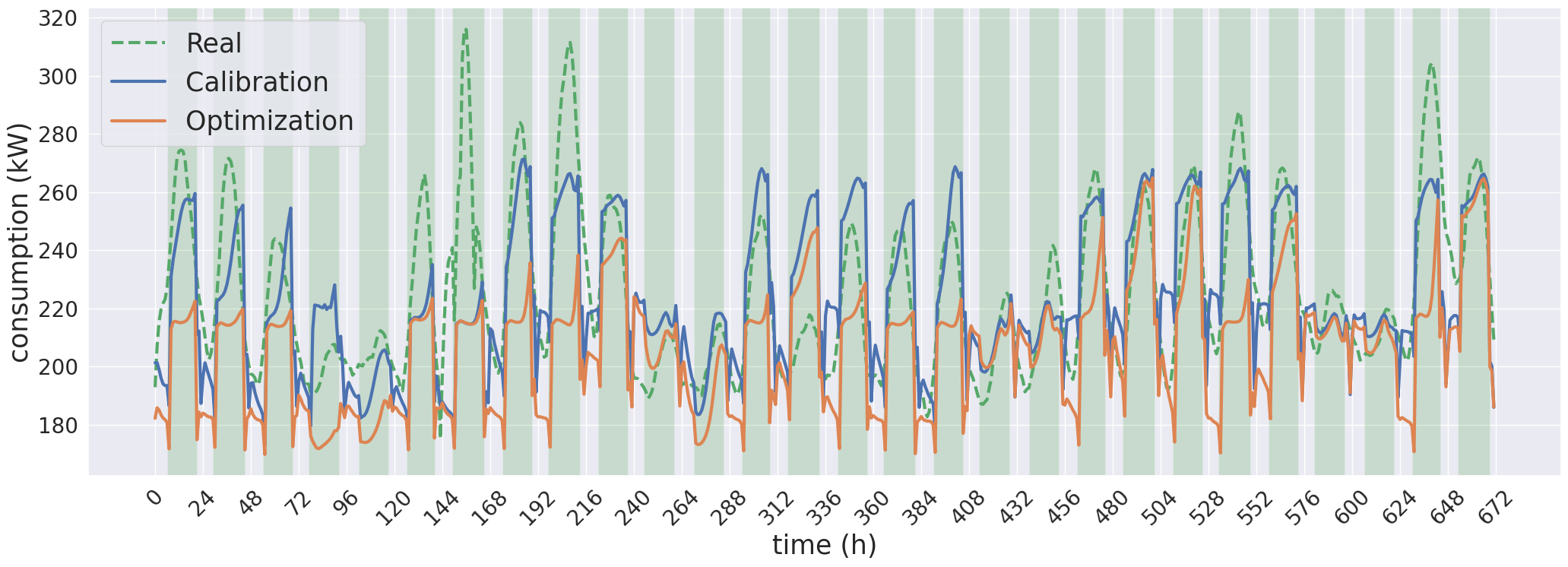}
	\includegraphics[width=0.95\textwidth]{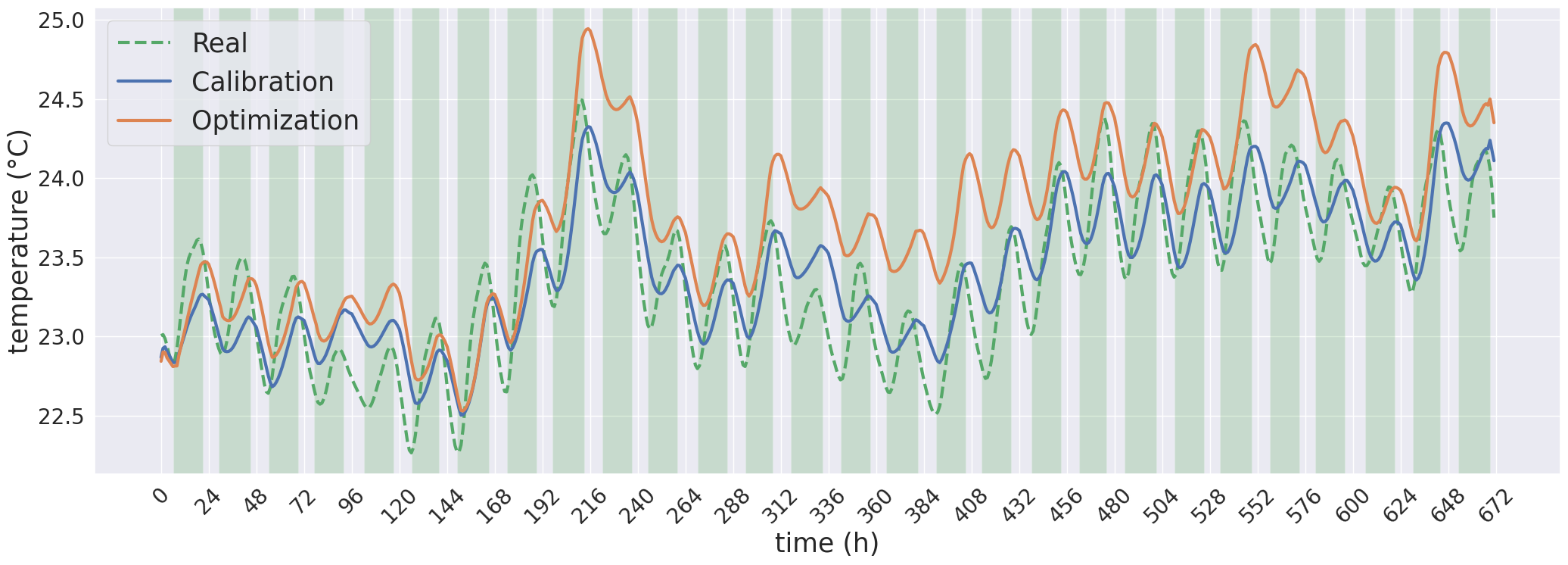}
	\caption{Consumption and temperature simulations after optimization (metamodel) for the Stanley building.} 
	\label{fig:timeseriesafteroptim}
\end{figure}

The parameters $\psi_k$ associated with the HVAC system can then be optimized for a given set of weather data $\varphi_k$.
The optimization tasks consists in finding a set a usage related parameters that reduce consumption while keeping the same level of comfort.
Optimizing energy consumption requires minimizing two conflicting objectives, making it impossible to find a solution that optimize both objectives simultaneously.
Instead, we search for optimal compromises between energy consumption and comfort, and plot each proposition to form a Pareto front, see Figure~\ref{fig:pareto}.
Combinations of energy consumption and comfort are unreachable below the Pareto front, and suboptimal above; we always aim at sampling points at the intersection.
Indeed, for any such optimal compromise, we can always get a higher level of comfort, for the price of a higher consumption.
The consumption criteria is the energy load during the month; the comfort criteria is the gap between indoor temperature and a constant reference temperature $T^*$:
\begin{align*}
	\mathrm{Comf} = \frac{1}{N_{\mathrm{Occ}}}\left(\sum_{k=1}^{N_{\mathrm{Occ}}} \mathds{1}_{k\in \mathrm{Occ}}(\widehat T_k - T^*)^2\right)^{1/2}\quad\mathrm{and} & \quad
	\bar Q = \frac{1}{N} \sum_{k=1}^{N} \widehat Q_k\,,
\end{align*}
where $T^*=22.5^{\circ}C$, $N^{\mathrm{opt}}$ is the number of hours to be considered in the optimization process and $\mathrm{Occ}$ is a subset of daytime hours specifying at which hours the target temperature has to be reached in the building.
%
Following recent works in building energy optimization, we search for a set of optimal parameters using NSGA-II, see \cite{Deb2000AFE}, another evolutionary algorithm, but adapted to multi objective problems.
An implementation can be found in the Pygmo\footnote{https://esa.github.io/pygmo2/} library.
In the absence of a stopping condition, we simply run the optimization for 3000 iterations (2 hours).
Results can be viewed as a Pareto front which is given in Figure \ref{fig:pareto} for the second month used in the calibration process.
As observed during calibration, this process can take a colossal number of iterations before achieving satisfactory results, once again justifying the use of a much faster metamodel.
The predicted time series associated with the BMS parameters selected in Figure~\ref{fig:pareto} are given in Figure~\ref{fig:timeseriesafteroptim}.
The relative gain, as well as the expected energy savings for both building are available in Table~\ref{tab:optim}.

\begin{table}[htpb]
	\centering
	\caption{Energy gain after optimization. Relative gain represents the energy load reduction between calibration and optimization steps, when maintaining the initial level of comfort. We then apply this coefficient to the real monthly consumption to obtain the reduction forecast in MWh. We also provide a more interesting reduction obtained by reducing the comfort criteria by $0.5^\circ\!$C.}
	\label{tab:optim}
	\resizebox{\textwidth}{!}{
		\begin{tabular}{c|cc|cc}
				    & relative gain (\%) & prevision (MWh) & relative gain / $0.5^\circ\!$ C (\%) & prevision / $0.5^\circ\!$ C (MWh) \\
			\hline
			Stanley     & 5.32               & 8.05            & 10.5                                 & 15.9                              \\
			Livingstone & 9.92               & 9.96            & 17.3                                 & 17.3
		\end{tabular}
	}
\end{table}

\section{Conclusion}
Optimizing building energy consumption is a challenging task which requires carefully integrated sensors to solve a multi-objective problem based on computationally very intensive calibration and optimization procedures.
For this reason, most related works focus on a single aspect of the problem.
In this paper, we proposed an end-to-end meta modelling methodology to tackle this problem in weakly instrumented buildings, with a small computational budget.

Our methodology relies on samples from a simulation program, such as TRNSYS, wherein building behavior can be approximated by an equivalent model defined by energy managers.
We experimented with various deep learning architectures to substitute the physical simulator, such as RNN and Transformer, and report sensible improvements over models described in the literature.
Additionally, because these models allow the computation of multiple simulations in parallel, we were able to significantly cut the computation time of the calibration and optimization steps, by a factor of over 70.

We implemented a calibration step in order to estimate specific features of a building, such as heat capacity, window to wall ratio, or air infiltration rates.
Although the metamodel approximates a simplified equivalent model, we successfully calibrated two real buildings based on data gathered from only a few sensors.
Convergence was reached for the first building after $300$ iterations, and displayed encouraging results when we applied estimated parameters to predict fresh data from unseen weeks.
Here, we demonstrated that calibration with TRNSYS achieved similar results, although requiring a much longer computation time.
On the second building, it took about $10,000$ iterations before the metamodel reached convergence, which would have represented $100$ computation days with TRNSYS.
This experiment showed that by reducing computation time, the metamodel is able to calibrate buildings which would have been impossible for physical simulators.
We believe that the greater difficulty to calibrate some buildings can be explained by less informative data due to noisy sensors, change in the building's usage during the experiment, etc.

Lastly, we were able to reduce energy consumption while preserving the same level of comfort in the building.
We explored the space of possible usage settings and reached a family of solutions offering equivalent compromises of both objectives.
For the solution providing the closest level of comfort as the historical data, we obtained a consumption reduced by 5\% and 10\%.
By depreciating the comfort objective of $0.5^\circ\!$C, we were able to further reduce consumption by up to $17\%$.

Our study highlights the potential for hourly optimization of multiple buildings, from a single metamodel training.
Future works could focus on exploring a wider variety of buildings, to assess the reach of adaptability and applicability of such a metamodel.
Because the training dataset was sampled based on a simplified equivalent model, calibration performances may drop when facing more complex buildings. 
An open question is the design of an automatic clustering procedure of many buildings from the equivalent description provided by energy managers. Such unsupervised clustering would allow the design of few metamodels trained only once and specifically built to target all buildings in each cluster.

Integration of a metamodel in the end-to-end pipeline could also open the way for improving calibration performances.
Indeed, we believe that one of the main obstacles in calibrating lies in the observation noise of sensors data.
While physical simulators can only simulate well defined thermal diffusion equations, we can modify our metamodel to take into account this inherent noise. We believe that this leads the way to interesting future research works for instance on the development on new finetuning procedures. Here, real data obtained in a building would be used to update the weights of the metamodel in order to capture specific behaviors that are not described by the physical simulator. Such approaches could benefit from statistical models such as general state space models to capture the complex statistical structure of the observations.


\bibliographystyle{apalike}
\bibliography{references}

\clearpage
\appendix

\section{Additional illustrations}
\begin{figure}[htpb]
        \centering
        \includegraphics[width=.9\linewidth]{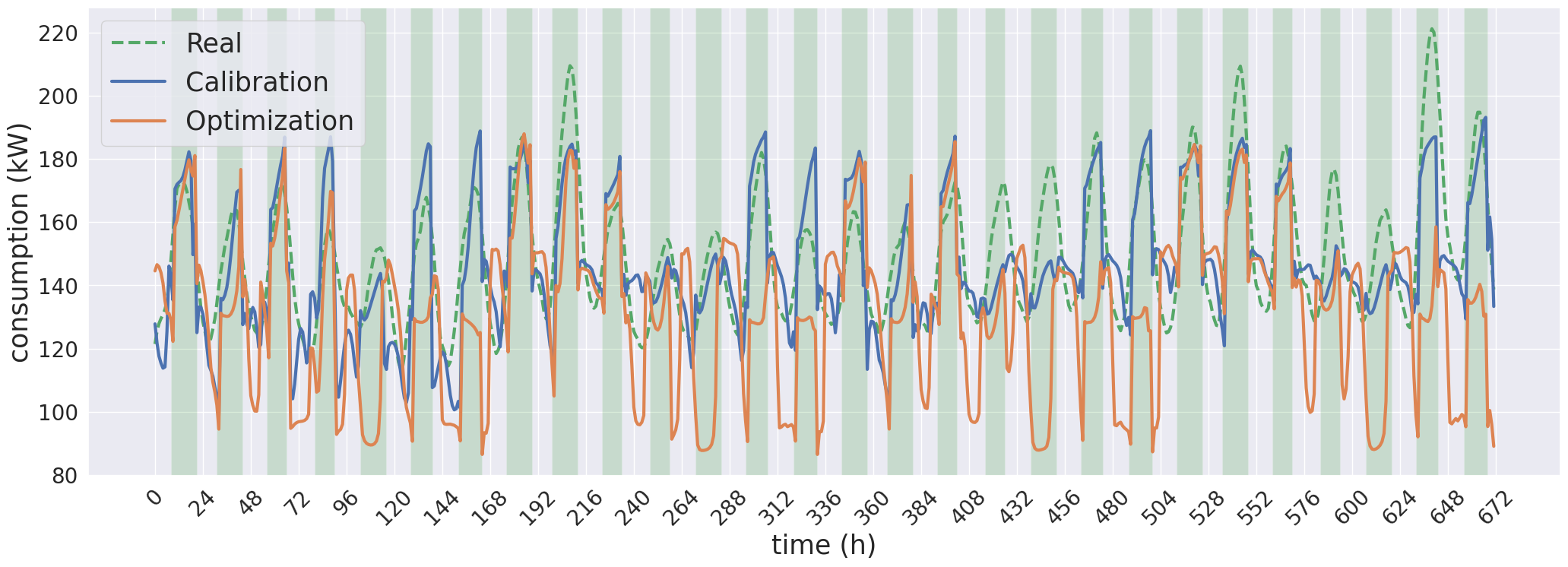}
        \includegraphics[width=.9\linewidth]{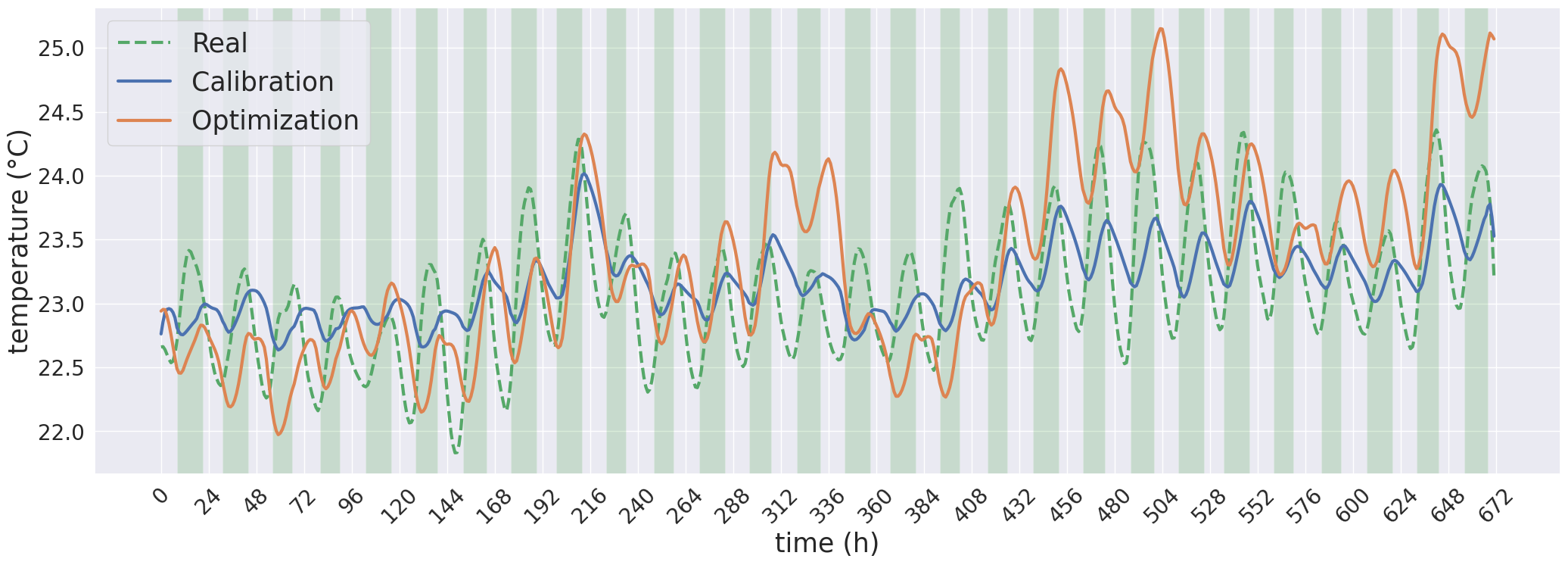}
        \caption{Indoor temperature and consumption for Real data, Calibration and Optimization for Livingstone.}%
\end{figure}

\begin{figure}[htpb]
        \centering
        \includegraphics[width=\linewidth]{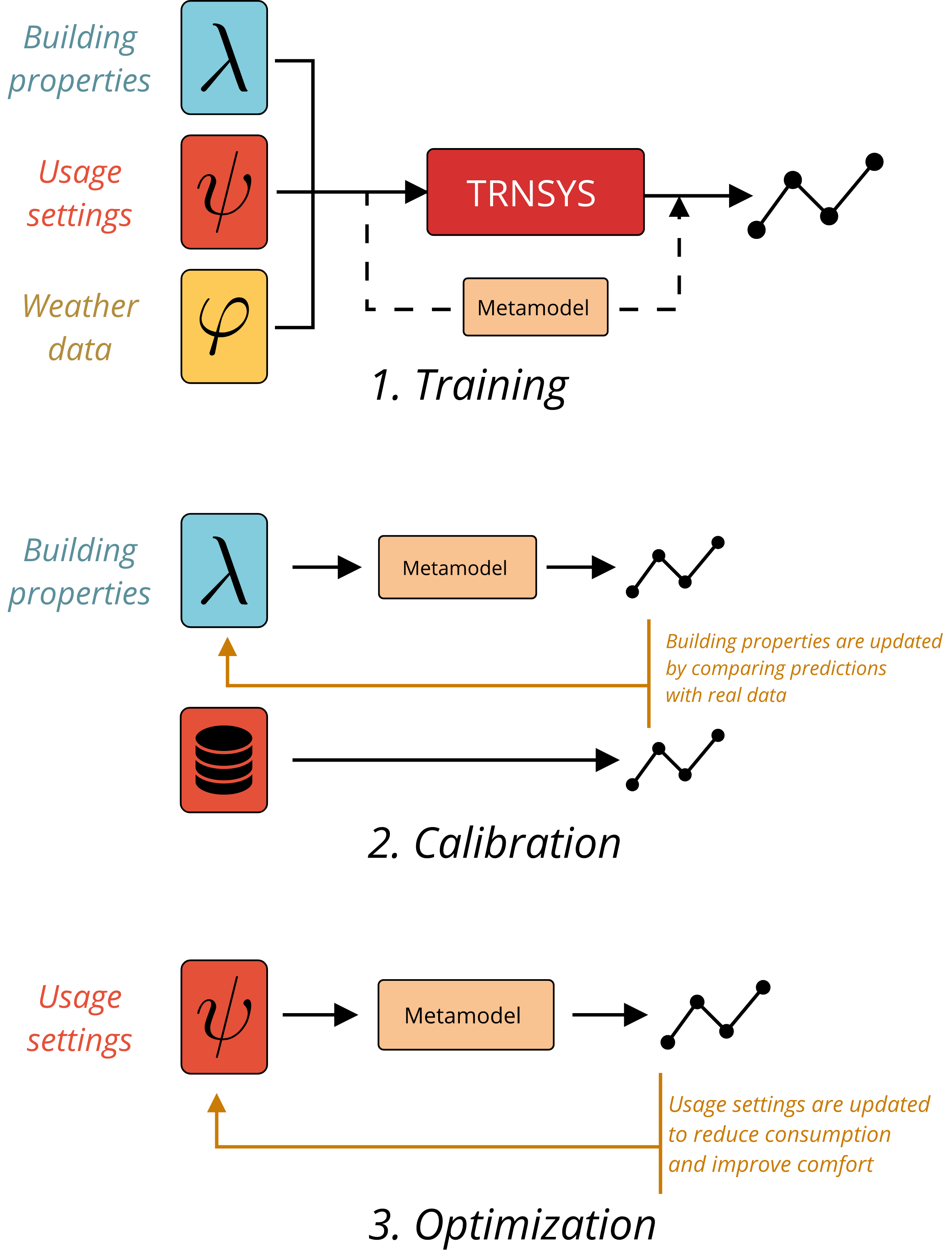}
        \caption{A flowchart summarizing our end-to-end methodology.}
        \label{fig:flowchart}
\end{figure}

\clearpage
\newpage

\section{Ranges used to train the metamodel}
\begin{table}[ht]
        \centering
        \resizebox{\textwidth}{!}{%
                \begin{tabular}{cccc}
                        Variable                                                                        & Minimum & Maximum & Step \\
                        \hline
                        airchange\_infiltration\_vol\_per\_h ($\mathrm{m}^3\mathrm{h}^{-1}$)            & 0.1     & 0.5     & 0.1  \\
                        capacitance\_kJ\_perdegreK\_perm3 ($\mathrm{kJ}\mathrm{K}^{-1}\mathrm{m}^{-3}$) & 50      & 300     & 10   \\
                        power\_VCV\_kW\_heat ($\mathrm{kW}$)                                            & 0       & 1000    & 100  \\
                        power\_VCV\_kW\_clim ($\mathrm{kW}$)                                            & 0       & 1000    & 100  \\
                        nb\_occupants                                                                   & 1000    & 2000    & 200  \\
                        nb\_PCs                                                                         & 1000    & 2000    & 200  \\
                        percent\_light\_night                                                           & 0       & 70      & 10   \\
                        percent\_PCs\_night                                                             & 0       & 70      & 10   \\
                        facade\_1\_thickness\_2 ($\mathrm{m}$)                                          & 0.05    & 0.15    & 0.05 \\
                        facade\_2\_thickness\_2 ($\mathrm{m}$)                                          & 0.05    & 0.15    & 0.05 \\
                        facade\_3\_thickness\_2 ($\mathrm{m}$)                                          & 0.05    & 0.15    & 0.05 \\
                        facade\_4\_thickness\_2 ($\mathrm{m}$)                                          & 0.05    & 0.15    & 0.05 \\
                        roof\_1\_thickness\_3 ($\mathrm{m}$)                                            & 0.05    & 0.15    & 0.05 \\
                        facade\_1\_window\_area\_percent                                                & 40      & 50      & 5    \\
                        facade\_2\_window\_area\_percent                                                & 40      & 50      & 5    \\
                        facade\_3\_window\_area\_percent                                                & 40      & 50      & 5    \\
                        facade\_4\_window\_area\_percent                                                & 40      & 50      & 5    \\
                        \hline
                        start\_occupation\_monday (h)                                                   & 7       & 9       & 1    \\
                        start\_occupation\_tuesday (h)                                                  & 7       & 9       & 1    \\
                        start\_occupation\_wednesday (h)                                                & 7       & 9       & 1    \\
                        start\_occupation\_thursday (h)                                                 & 7       & 9       & 1    \\
                        start\_occupation\_friday  (h)                                                  & 7       & 9       & 1    \\
                        end\_occupation\_monday (h)                                                     & 17      & 20      & 1    \\
                        end\_occupation\_tuesday (h)                                                    & 17      & 20      & 1    \\
                        end\_occupation\_wednesday (h)                                                  & 17      & 20      & 1    \\
                        end\_occupation\_thursday  (h)                                                  & 17      & 20      & 1    \\
                        end\_occupation\_friday  (h)                                                    & 17      & 20      & 1    \\
                        \hline                                                                                                \\
                \end{tabular}}
        \caption{List of parameters contained in $\lambda$, along with sampling and calibration ranges. During training of the metamodel, occupation values are converted in a one dimensional time serie, with value $0$ or $1$ based on the occupation state of the building.}
        \label{tab:lambda}
\end{table}

\begin{table}[ht]
        \centering
        \begin{tabular}{cccc}
                Variable                                 & Minimum & Maximum & Step \\
                \hline
                start\_clim\_day (h)                     & 7       & 9       & 1    \\
                end\_clim\_day (h)                       & 18      & 20      & 1    \\
                t\_clim\_red\_day ($^\circ\mathrm{C}$)   & 24      & 30      & 0.5  \\
                t\_clim\_conf\_day ($^\circ\mathrm{C}$)  & 20      & 24      & 0.5  \\
                start\_heat\_day (h)                     & 6       & 8       & 1    \\
                end\_heat\_day (h)                       & 17      & 19      & 1    \\
                t\_heat\_red\_day ($^\circ\mathrm{C}$)   & 17      & 22      & 0.5  \\
                t\_heat\_conf\_day ($^\circ\mathrm{C}$)  & 22      & 24      & 0.5  \\
                start\_ventilation\_day (h)              & 7       & 9       & 1    \\
                end\_ventilation\_day (h)                & 18      & 20      & 1    \\
                t\_ventilation\_day ($^\circ\mathrm{C}$) & 18      & 26      & 0.5  \\
                vol\_ventilation\_day                    & 0.7     & 1.7     & 0.3  \\
                \hline                                                         \\
        \end{tabular}
        \caption{List of variables contained in $\psi_k$, along with their ranges. Each parameter can hold a different value for each day of the week. For ease of reading, we replaced them by a single line, as the ranges are the same for every day.}
        \label{tab:psi}
\end{table}

\begin{table}[ht]
        \centering
        \begin{tabular}{*2c}
                Variable & Description                   \\
                \hline
                DNI      & Direct Normal Irradiance      \\
                IBEAM\_H & Direct Horizontal Irradiance  \\
                IBEAM\_N & Direct Normal Irradiance      \\
                IDIFF\_H & Diffuse Horizontal Irradiance \\
                IGLOB\_H & Global Horizontal Irradiance  \\
                RHUM     & Outdoor Relative Humidity     \\
                TAMB     & Outdoor temperature           \\
                \hline                              \\
        \end{tabular}
        \caption{Weather data as contained in $\varphi_k$.}
        \label{tab:phi}
\end{table}

\begin{table}[ht]
    \centering
    \resizebox{\textwidth}{!}{%
    \begin{tabular}{cc}
    Variable & Description\\
    \hline
	Q\_AC\_OFFICE & AC consumption \\
	Q\_HEAT\_OFFICE & Heat consumption \\
	Q\_PEOPLE & Heating power due to human activities in the building \\
	Q\_EQP & Consumption of equipment, such as computers, elevators, fridges \\
	Q\_LIGHT & Consumption of lights \\
	Q\_AHU\_C & Consumption of AHU when cooling outside air \\
	Q\_AHU\_H & Consumption of AHU when heating outside air \\
	T\_INT\_OFFICE & Indoor temperature\\
    \hline \\
    \end{tabular}}
    \caption{Output variables of the equivalent model designed by the energy managers, contained in $Y_k$.}
    \label{tab:Yk}

\end{table}

\end{document}